\documentclass[a4paper]{article}

\usepackage{INTERSPEECH2019}
\usepackage{comment,cite,url}
\usepackage{color}
\usepackage{boldline}

\newcommand{\refeq}[1]{(\ref{eq:#1})}
\newcommand{\refeqs}[2]{(\ref{eq:#1}) and (\ref{eq:#2})}

\newcommand{\refsubsec}[1]{\ref{subsec:#1}}

\newcommand{\reffig}[1]{Fig. \ref{fig:#1}}

\def\Vec#1{\boldsymbol{\mathbf{#1}}}

\def\x{\Vec{x}}

\def\y{\Vec{y}}
\def\c{\Vec{c}}

\def\encdis{q_{\phi}}
\def\decdis{p_{\theta}}

\def\CN{{\mathcal N}}

\def\z{\Vec{z}}

\def\0{{\mathbf 0}}
\def\I{{\mathbf I}}
\def\vmu{{\boldsymbol \mu}}
\def\vsigma{{\boldsymbol \sigma}}
\def\vepsilon{{\boldsymbol \epsilon}}
\def\phia{\phi_{\rm a}}
\def\phiv{\phi_{\rm v}}
\def\thetaa{\theta_{\rm a}}
\def\thetav{\theta_{\rm v}}

\allowdisplaybreaks

\title{
Crossmodal Voice Conversion
}
\name{Hirokazu Kameoka, Kou Tanaka, Aar\'on Valero Puche, Yasunori Ohishi, Takuhiro Kaneko}
\address{
  NTT Communication Science Laboratories, Nippon Telegraph and Telephone Corporation
  }
\email{hirokazu.kameoka.uh@hco.ntt.co.jp}

\begin{document}

\maketitle
\begin{abstract}
Humans are able to imagine a person's voice from the person's appearance and imagine the person's appearance from his/her voice. In this paper, we make the first attempt to develop a method that can convert speech into a voice that matches an input face image and generate a face image that matches the voice of the input speech by leveraging the correlation between faces and voices. We propose a model, consisting of a speech converter, a face encoder/decoder and a voice encoder. We use the latent code of an input face image encoded by the face encoder as the auxiliary input into the speech converter and train the speech converter so that the original latent code can be recovered from the generated speech by the voice encoder. We also train the face decoder along with the face encoder to ensure that the latent code will contain sufficient information to reconstruct the input face image. We confirmed experimentally that a speech converter trained in this way was able to convert input speech into a voice that matched an input face image and that the voice encoder and face decoder can be used to generate a face image that matches the voice of the input speech.
\end{abstract}

\section{Introduction}

Humans are able to imagine a person's voice solely from that person's appearance 
and imagine the person's appearance solely from his/her voice. 
Although such predictions are not always accurate, 
the fact that 
we can sense if there is a mismatch between 
voice and appearance
should indicate the possibility of being a certain correlation between
voices and appearance. 
In fact, recent studies by Smith et al. \cite{Smith2016} have revealed that 
the information provided by faces
and voices is so similar that people can match novel faces and
voices of the same sex, ethnicity, and age-group at a level
significantly above chance.
Here, an interesting question is whether 
it is technically possible to 
predict the voice of a person only from an image of his/her face
and predict a person's face only from his/her voice.  
In this paper, we make the first attempt to develop a method that can
convert speech into a voice that matches 
an input face image and that can generate a face image that matches  
the voice providing input speech by learning and leveraging the underlying 
correlation between faces and voices. 

Several attempts have recently been made to tackle
the tasks of crossmodal audio/image processing, including 
voice/face recognition \cite{Nagrani2018} and
audio/image generation \cite{Chen2017,Zhou2018,Hao2018}.
The former task involves detecting which of two given face images is that of the speaker,
given only an audio clip of someone speaking.
Hence, this task differs from ours in that it
does not involve audio/image generation. 
The latter task involves generating sounds from images/videos.
The methods presented in \cite{Chen2017,Zhou2018,Hao2018} are designed to predict very short sound clips (e.g., 0.5 to 2 seconds long) such as the sounds made by musical instruments, dogs, and babies crying, 
and are unsuited to generating longer audio clips with richer variations in time such as speech utterances.
By contrast, our task is crossmodal voice conversion (VC), namely converting given speech utterances
where the target voice characteristics are determined by visual inputs. 

VC is a technique for converting 
the voice characteristics of an input utterance 
such as the perceived identity of a speaker 
while preserving linguistic information. 
Potential applications of VC techniques include 
speaker-identity modification, 
speaking aids, speech enhancement, and pronunciation conversion.
Typically, many conventional VC methods utilize
accurately aligned parallel utterances of source and target speech
to train acoustic models for feature mapping \cite{Stylianou1998,Toda2007,Kobayashi2018}.
Recently, some attempts have also been made to develop non-parallel VC methods \cite{Xie2016,Kinnunen2017,Hsu2016,Hsu2017,Saito2018,Kameoka2018b,Kaneko2017,Kameoka2018a}, 
which require no parallel utterances, transcriptions, or time alignment procedures.
One approach to non-parallel VC involves a framework based on conditional variational autoencoders (CVAEs) 
\cite{Hsu2016,Hsu2017,Saito2018,Kameoka2018b}.
As the name implies, variational autoencoders (VAEs) \cite{Kingma2014a} are a probabilistic counterpart of autoencoders, 
consisting of encoder and decoder networks. CVAEs \cite{Kingma2014b} are an extended
version of VAEs where the encoder and decoder networks can additionally take an auxiliary input. 
By using acoustic features as the training examples and the associated attribute 
(e.g., speaker identity) labels as the auxiliary input, 
the networks are able to learn how to convert 
an attribute of source speech to a target attribute according
to the attribute label fed into the decoder.
As a different approach, in \cite{Kaneko2017} we proposed
a method using a variant of a generative adversarial network (GAN) \cite{Goodfellow2014} called a cycle-consistent GAN (CycleGAN) 
\cite{Zhu2017,Kim2017,Yi2017}. 
Although this method was shown to work reasonably well, one major limitation is that
it is designed to learn only mappings between a pair of domains.
To overcome this limitation, we subsequently proposed in
\cite{Kameoka2018a} a method incorporating an extension of
CycleGAN called StarGAN \cite{Choi2017}. This method is capable of
simultaneously learning mappings between multiple domains
using a single generator network where the attributes of the
generator outputs are controlled by an auxiliary input. 
StarGAN uses an auxiliary classifier to train the generator
so that the attributes of the
generator outputs are correctly predicted by the classifier.
We further proposed a method based on a concept that combined StarGAN and CVAE, 
called an auxiliary classifier VAE (ACVAE) \cite{Kameoka2018b}. 
An ACVAE employs a generator with a CVAE structure   
and uses an auxiliary classifier to train the generator in the same way as StarGAN.
Training the generator in this way can be interpreted as increasing the lower bound of 
the mutual information between the auxiliary input and the generator output.

In this paper, we propose extending the idea behind the ACVAE
to build a model for crossmodal VC. 
Specifically, we use the latent code of an auxiliary face image input 
encoded by a face encoder as 
the auxiliary input into the speech generator 
and use a voice encoder to train the generator
so that the original latent code can be recovered from 
the generated speech using the voice encoder. 
We also train a face decoder along with the face encoder to ensure that 
the latent code will contain sufficient information to 
reconstruct the input face image. 
In this way, 
the speech generator is expected to
learn how to convert input speech into a voice characteristic that matches 
an auxiliary face image input 
and the voice encoder and the face decoder 
can be used to generate a face image that matches the voice characteristic of input speech.

\section{Method}

\subsection{Variational Autoencoder (VAE)}
\label{subsec:vae}

Our model employs VAEs \cite{Kingma2014a,Kingma2014b} as building blocks.
Here, we briefly introduce the principle behind VAEs.

VAEs are stochastic neural network models consisting of encoder and decoder networks.
The encoder aims to encode given data $\x$ into
a (typically) lower dimensional latent representation $\z$
whereas the decoder aims to recover the data $\x$ from the latent representation $\z$. 
The decoder is modeled as a neural network (decoder network) that 
produces a set of parameters for a conditional distribution $\decdis(\x|\z)$
where $\theta$ denotes the network parameters.
To obtain an encoder using $\decdis(\x|\z)$, we must 
compute the posterior $\decdis(\z|\x) = \decdis(\x|\z)p(\z)/\decdis(\x)$.
However, computing the exact posterior is usually difficult 
since $\decdis(\x)$ involves an intractable integral over $\z$.
The idea of VAEs is to sidestep the direct computation of this posterior by 
introducing another neural network (encoder network) 
for approximating the exact posterior $\decdis(\z|\x)$. 
As with the decoder network, the encoder network generates a set of parameters for the conditional distribution $\encdis(\z|\x)$ where $\phi$ denotes the network parameters. 
The goal of VAEs is to learn the parameters of 
the encoder and decoder networks so that 
the encoder distribution $\encdis(\z|\x)$ becomes consistent with the posterior $\decdis(\z|\x)\propto \decdis(\x|\z)p(\z)$. 
We can show that
the Kullback-Leibler (KL) divergence 
between $\encdis(\z|\x)$ and $\decdis(\z|\x)$
is given as
\begin{multline}
{\rm KL}[\encdis(\z|\x) \| \decdis(\z|\x)] = \log p(\x)\\
- 
\mathbb{E}_{\z\sim \encdis(\z|\x)}[\log \decdis(\x|\z)] 
+ {\rm KL}[ \encdis(\z|\x) \| p(\z) ].
\end{multline}
Here, it should be noted that 
since ${\rm KL}[\encdis(\z|\x) \| \decdis(\z|\x)] \ge 0$, 
$\mathbb{E}_{\z\sim \encdis(\z|\x)}[\log \decdis(\x|\z)] - {\rm KL}[ \encdis(\z|\x) \| p(\z) ]$ is shown to be a lower bound for $\log p(\x)$.
Given training examples,
\begin{multline}
\mathcal{J}(\theta,\phi) = 
\mathbb{E}_{\x\sim p_{\rm d}(\x)}
\big[
\mathbb{E}_{\z\sim \encdis(\z|\x)}[\log \decdis(\x|\z)] 
\\
- {\rm KL}[ \encdis(\z|\x) \| p(\z) ]
\big],
\label{eq:VAEloss}
\end{multline}
can be used as the training criterion to be maximized with respect to 
$\theta$ and $\phi$,
where $\mathbb{E}_{\x\sim p_{\rm d}(\x)}[\cdot]$ denotes the sample mean over the training examples.
Obviously, $\mathcal{J}(\theta,\phi)$ is maximized when 
the exact posterior is obtained
$\encdis(\z|\x) = \decdis(\z|\x)$.

One typical way of modeling $\encdis(\z|\x)$, $\decdis(\x|\z)$ and $p(\z)$ is to assume
Gaussian distributions
\begin{align}
\encdis(\z|\x) &= \mathcal{N}(\z|\vmu_{\phi}(\x), {\rm diag}(\vsigma_{\phi}^2(\x))),
\label{eq:q(z|s)}
\\
\decdis(\x|\z) &= \mathcal{N}(\x|\vmu_{\theta}(\z), {\rm diag}(\vsigma_{\theta}^2(\z))),
\label{eq:p(s|z)}
\\
p(\z) &= \mathcal{N}(\z|{\bf 0},{\bf I}),
\label{eq:p(z)}
\end{align}
where $\vmu_{\phi}(\x)$ and $\vsigma_{\phi}^2(\x)$ are the outputs of an encoder network with parameter $\phi$,
and $\vmu_{\theta}(\z)$ and $\vsigma_{\theta}^2(\z)$ are the outputs of a decoder network with parameter $\theta$.
The first term of \refeq{VAEloss} 
can be interpreted as an autoencoder reconstruction error.
Here, it should be noted that to compute 
this term, we must compute the 
expectation with respect to $\z\sim \encdis(\z|\x)$. 
Since this expectation cannot be expressed in an analytical form,
one way of computing it involves using a Monte Carlo approximation.
However, simply sampling $\z$ from $\encdis(\z|\x)$ does not work,
since 
once $\z$ is sampled, $\z$ is no longer a function of $\phi$
and so it becomes impossible to evaluate
the gradient of $\mathcal{J}(\theta,\phi)$ with respect to $\phi$.
Fortunately, 
by using a reparameterization 
$\z = \vmu_{\phi}(\x) + \vsigma_{\phi}(\x) \odot \vepsilon$ with 
$\vepsilon \sim \mathcal{N}(\vepsilon|\0,\I)$, 
sampling $\z$ from $\encdis(\z|\x)$ can be replaced by
sampling $\vepsilon$ 
from the distribution, which is independent of $\phi$.
This allows us to compute the gradient of the first term of 
$\mathcal{J}(\theta,\phi)$ 
with respect to $\phi$
by using a Monte Carlo approximation of the expectation $\mathbb{E}_{\z\sim\encdis(\z|\x)}[\cdot]$. 
The second term is given as the negative KL divergence 
between $\encdis(\z|\x)$ and $p(\z)=\mathcal{N}(\z|\0,\I)$. 
This term can be interpreted as a regularization term that 
forces each element of the encoder output to be uncorrelated and normally distributed.
It should be noted that
when $\encdis(\z|\x)$ and $p(\z)$ are Gaussians, 
this term can be expressed as a function of $\phi$.

Conditional VAEs (CVAEs) \cite{Kingma2014b} are an extended version of VAEs 
with the only difference being that
the encoder and decoder networks can take an auxiliary input 
$c$. With CVAEs, \refeqs{q(z|s)}{p(s|z)} are replaced with
\begin{align}
\encdis(\z|\x,c) &= \mathcal{N}(\z|\vmu_{\phi}(\x,c), {\rm diag}(\vsigma_{\phi}^2(\x,c))),
\label{eq:q(z|s,c)}
\\
\decdis(\x|\z,c) &=
\mathcal{N}(\x|\vmu_{\theta}(\z,c), {\rm diag}(\vsigma_{\theta}^2(\z,c))),
\label{eq:p(s|z,c)}
\end{align}
and the training criterion to be maximized becomes
\begin{multline}
\mathcal{J}(\theta,\phi) 
=
\mathbb{E}_{
(\x,c)\sim p_{\rm d}(\x,c)
}\big[ 
\mathbb{E}_{\z\sim \encdis(\z|\x,c)}[\log \decdis(\x|\z,c)] 
\\
- {\rm KL}[ \encdis(\z|\x,c) \| p(\z) ]
\big],
\label{eq:CVAEloss}
\end{multline}
where $\mathbb{E}_{(\x,c)\sim p_{\rm d}(\x,c)}[\cdot]$ denotes the sample mean over the training examples. 

\subsection{Proposed model}
\label{subsec:proposed}

\begin{figure}[t!]
\centering
  \centerline{\includegraphics[width=.5\linewidth]{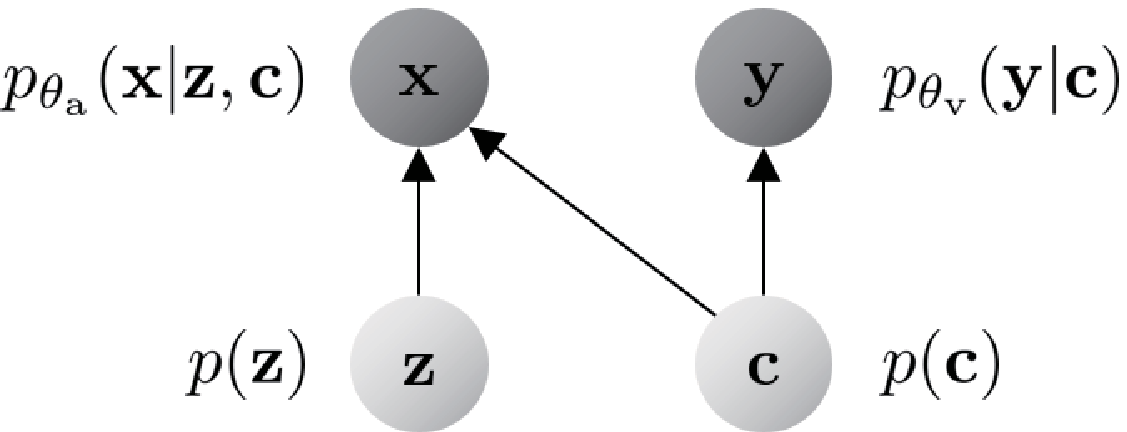}}
  \vspace{-0ex}
  \caption{Graphical model for $p(\x,\y)$}
\label{fig:GraphicalModel}
\end{figure}

We use $\x = [\x_1,\ldots,\x_N] \in\mathbb{R}^{D\times N}$ and $\y\in \mathbb{R}^{I\times J}$ to denote
the acoustic feature vector sequence of a speech utterance and the face image of the corresponding speaker.
Now, we combine two VAEs to model 
the joint distribution of $\x$ and $\y$. 
The encoder for speech (hereafter, the {\bf utterance encoder}) aims to 
encode $\x$ into a time-dependent latent variable sequence $\z = [\z_1,\ldots,\z_{N'}] \in\mathbb{R}^{D'\times N'}$ 
whereas the decoder (hereafter, the {\bf utterance decoder}) 
aims to reconstruct $\x$ from $\z$ using an auxiliary input $\c$.
Ideally, we would like $\z$ to capture only the linguistic information
contained in $\x$ and $\c$ to contain information about the target voice characteristics. 
Hence, we expect that the encoder and decoder  
work as acoustic models for speech recognition and speech synthesis
so that they can be used to convert the voice of an input utterance 
according to the auxiliary input $\c$. 
We use the time-independent latent code of an image $\y$ encoded by 
the encoder for face images (hereafter, the {\bf face encoder}) 
as the auxiliary input $\c$ into the utterance decoder. 
The decoder for face images (hereafter, the {\bf face decoder}) is designed to 
reconstruct $\y$ from $\c$.
\reffig{GraphicalModel} shows the assumed graphical model for the joint distribution $p(\x,\y)$.

Our model can be formally described as follows.
The utterance/face decoders 
and the utterance/face encoders
are represented 
as the conditional distributions $p_{\thetaa}(\x|\z,\c)$,
$p_{\thetav}(\y|\c)$, $q_{\phia}(\z|\x)$ and $q_{\phiv}(\c|\y)$,
expressed using NNs with parameters $\thetaa$, $\thetav$, $\phia$ and $\phiv$,
respectively.
Our aim is to approximate the exact posterior
$p(\z,\c|\x,\y) \propto p_{\thetaa}(\x|\z,\c) p_{\thetav}(\y|\c)$
by $q(\z,\c|\x,\y) = q_{\phia}(\z|\x) q_{\phiv}(\c|\y)$.
The KL divergence between these distributions
is given as 
\begin{align}
&{\rm KL}[q(\z,\c|\x,\y)\|p(\z,\c|\x,\y)] 
= \log p(\x,\y) \nonumber\\
&~~~~~~~~
- 
\mathbb{E}_{
\c\sim q_{\phiv}(\c|\y),
\z\sim q_{\phia}(\z|\x)
}
[\log p_{\thetaa}(\x|\z,\c)]
\nonumber\\
&~~~~~~~~
-
\mathbb{E}_{
\c\sim q_{\phiv}(\c|\y)
}
[\log p_{\thetav}(\y|\c)]
\nonumber\\
&~~~~~~~~
+{\rm KL}[ q_{\phia}(\z|\x) \| p(\z) ]
+{\rm KL}[ q_{\phiv}(\c|\y) \| p(\c) ].
\end{align}
Hence, 
given the training examples of speech and face pairs
$\{\x_m,\y_m\}_{m=1}^{M}$, 
we can use 
\begin{align}
&\mathcal{J}(\thetaa,\phia,\thetav,\phiv) \nonumber\\
=&
\mathbb{E}_{
(\x,\y) \sim p_{\rm d}(\x,\y) 
}
\mathbb{E}_{
\c\sim q_{\phiv}(\c|\y),\z \sim q_{\phia}(\z|\x)
}
[\log p_{\thetaa}(\x|\z,\c)] 
\nonumber\\
+&
\mathbb{E}_{
\y \sim p_{\rm d}(\y)
}
\mathbb{E}_{\c \sim q_{\phiv}(\c|\y)}
[\log p_{\thetav}(\y|\c)] 
\nonumber\\
-&
\mathbb{E}_{
\x \sim p_{\rm d}(\x) 
}
{\rm KL}[ q_{\phia}(\z|\x) \| p(\z) ]
\nonumber\\
-&
\mathbb{E}_{
\y \sim p_{\rm d}(\y)
}
{\rm KL}[ q_{\phiv}(\c|\y) \| p(\c) ],
\end{align}
as the training criterion to be maximized
with respect to $\thetaa$, $\phia$, $\thetav$, and $\phiv$,
where $\mathbb{E}_{({\x},{\y}) \sim p_{\rm d}({\x},{\y})}[\cdot]$,
$\mathbb{E}_{\x \sim p_{\rm d}({\x})}[\cdot]$
and
$\mathbb{E}_{\y \sim p_{\rm d}({\y})}[\cdot]$
denote the sample means over the training examples.
We assume the encoder/decoder distributions for $\x$ and $\y$ to be 
Gaussian distributions:
\begin{align}
q_{\phia}(\z|\x)&=\CN(\z|\vmu_{\phia}(\x),{\rm diag}(\vsigma_{\phia}^2(\x))),\\
p_{\thetaa}(\x|\z,\c)&=\CN(\x|\vmu_{\thetaa}(\z,\c),{\rm diag}(\vsigma_{\thetaa}^2(\z,\c))),\\
q_{\phiv}(\c|\y)&=\CN(\c|\vmu_{\phiv}(\y),{\rm diag}(\vsigma_{\phiv}^2(\y))),\\
p_{\thetav}(\y|\c)&=\CN(\y|\vmu_{\thetav}(\c),{\rm diag}(\vsigma_{\thetav}^2(\c))),
\end{align}
where 
$\vmu_{\phia}(\x)$ and $\vsigma_{\phia}^2(\x)$ are the outputs of the utterance encoder network,
$\vmu_{\thetaa}(\z,\c)$ and $\vsigma_{\thetaa}^2(\z,\c)$ are the outputs of the utterance decoder network,
$\vmu_{\phiv}(\y)$ and $\vsigma_{\phiv}^2(\y)$ are the outputs of the face encoder network,
and $\vmu_{\thetav}(\c)$ and $\vsigma_{\thetav}^2(\c)$ are the outputs of the face decoder network.
We further assume $p(\z)$ and $p(\c)$ to be standard Gaussian distributions, namely
$p(\z)=\CN(\z|\0,\I)$ and $p(\c)=\CN(\c|\0,\I)$. 
It should be noted that 
we can use the same reparametrization trick as in \refsubsec{vae} to 
compute the gradients of $\mathcal{J}(\thetaa,\phia,\thetav,\phiv)$
with respect to $\phia$ and $\phiv$.

Since there are no explicit restrictions on the manner in which the utterance decoder 
may use the auxiliary input $\c$, we introduce an information-theoretic regularization term
to assist the utterance decoder output to be correlated with $\c$ as far as possible. 
The mutual information for $\x\sim p_{\thetaa}(\x|\z,\c)$ and $\c$ 
conditioned on $\z$ 
can be written as
\begin{align}
\mathcal{I}(\thetaa) &= \iint p(\c',\x) \log \frac{p(\c',\x)}{p(\c')p(\x)} \mbox{d}\x\mbox{d}\c'\nonumber\\
&= \iint p(\x)p(\c'|\x) \log p(\c'|\x) \mbox{d}\x \mbox{d}\c' + H\nonumber\\
&= \mathbb{E}_{\x\sim p_{\thetaa}(\x|\z,\c),\c'\sim p(\c|\x)}[\log p(\c'|\x)] + H,
\label{eq:MI}
\end{align}
where $H$ represents the entropy of $\c$, which can be considered a constant term.
In practice, $\mathcal{I}(\thetaa)$ is hard to optimize directly since it requires
access to the posterior $p(\c|\x)$. Fortunately, we can obtain the lower bound of 
the first term of $\mathcal{I}(\thetaa)$ 
by introducing an auxiliary distribution $r(\c|\x)$
\begin{align}
&\mathbb{E}_{\x\sim p_{\thetaa}(\x|\z,\c),\c'\sim p(\c|\x)}[\log p(\c'|\x)]
\nonumber\\
=&\mathbb{E}_{\x\sim p_{\thetaa}(\x|\z,\c),\c'\sim p(\c|\x)}
\left[\log \frac{r(\c'|\x)p(\c'|\x)}{r(\c'|\x)}\right] 
\nonumber\\
\ge& \mathbb{E}_{\x\sim p_{\thetaa}(\x|\z,\c),\c'\sim p(\c|\x)}[\log r(\c'|\x)] 
\nonumber\\
=&
\mathbb{E}_{\x\sim p_{\thetaa}(\x|\z,\c)}[\log r(\c|\x)].
\label{eq:MI-LB}
\end{align}
This technique of lower bounding mutual information is called 
variational information maximization \cite{Barber2003short}. 
The equality holds in \refeq{MI-LB} when $r(\c|\x)=p(\c|\x)$.
Hence, maximizing the lower bound \refeq{MI-LB} with respect to $r(\c|\x)$ corresponds to 
approximating $p(\c|\x)$ by $r(\c|\x)$ as well as approximating $\mathcal{I}(\thetaa)$ by this lower bound.
We can therefore indirectly increase $\mathcal{I}(\thetaa)$ by increasing the lower bound alternately with respect to
$p_{\thetaa}(\x|\z,\c)$ and $r(\c|\x)$. One way to do this involves 
expressing $r(\c|\x)$ using an NN and training it along with all other networks. 
Let us use the notation $r_{\psi}(\c|\x)$ to indicate $r(\c|\x)$ expressed using an NN with parameter $\psi$. 
The role of $r_{\psi}(\c|\x)$ (hereafter, the {\bf voice encoder})
is to recover time-independent information about the voice characteristics of $\x$. 
For example, we can assume $r_{\psi}(\c|\x)$ to be a Gaussian distribution
\begin{align}
r_{\psi}(\c|\x) = 
\CN(\c|\vmu_{\psi}(\x),{\rm diag}(\vsigma_{\psi}^2(\x))),
\label{eq:r_psi}
\end{align}
where $\vmu_{\psi}(\x)$ and $\vsigma_{\psi}^2(\x)$ are the outputs 
of the voice encoder network. 
Under this assumption,
\refeq{MI-LB} becomes a negative weighted squared error between 
$\c\sim q_{\phiv}(\c|\y)$ and $\vmu_{\psi}(\x)$. Thus, 
maximizing \refeq{MI-LB} corresponds to forcing
the outputs of the face and voice encoders to be as consistent as possible.
Hence, the regularization term that we would like to maximize with respect to 
$\thetaa$, $\phia$,  $\phiv$ and $\psi$
becomes
\begin{multline}
\mathcal{R}(\thetaa,\phia,\phiv,\psi) 
= 
\mathbb{E}_{
\tilde{\x} \sim p_{\rm d}(\tilde{\x}),
\tilde{\y} \sim p_{\rm d}(\tilde{\y})
}
\\
\mathbb{E}_{
\z\sim q_{\phia}(\z|\tilde{\x}),
\c\sim q_{\phiv}(\c|\tilde{\y})}
\label{eq:R1}
\mathbb{E}_{
\x\sim p_{\thetaa}(\x|\z,\c)
}[\log r_{\psi}(\c|\x)],
\end{multline}
where 
$\mathbb{E}_{\tilde{\x} \sim p_{\rm d}(\tilde{\x})}[\cdot]$ 
and 
$\mathbb{E}_{\tilde{\y} \sim p_{\rm d}(\tilde{\y})}[\cdot]$ 
denote the sample means over the training examples.
Here, it should be noted
that to compute $\mathcal{R}(\thetaa,\phia,\phiv,\psi)$, 
we must sample $\z$ from $q_{\phia}(\z|\x)$, 
$\c$ from $q_{\phiv}(\c|\y)$ and $\x$ from $p_{\thetaa}(\x|\z,\c)$. 
Fortunately, we can use the same
reparameterization trick as in \refsubsec{vae} to compute the gradients
of $\mathcal{R}(\thetaa,\phia,\phiv,\psi)$ with respect to 
$\thetaa$, $\phia$, $\phiv$ and $\psi$.

\begin{figure*}[t!]
\centering
  \centerline{\includegraphics[width=.65\linewidth]{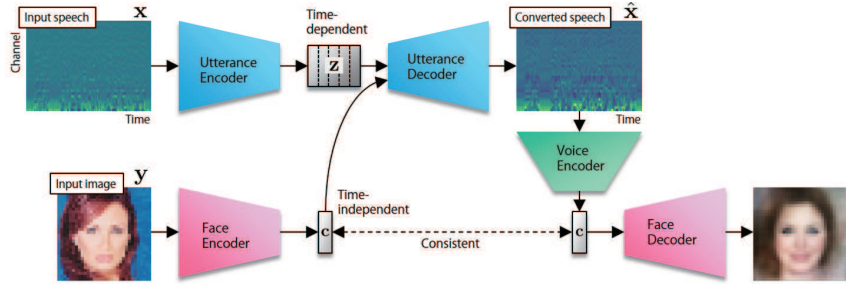}}
  \vspace{-0ex}
  \caption{Overview of the present model}
\label{fig:ModelOverview}
\end{figure*}

Overall, the training criterion to be maximized becomes
\begin{align}
\mathcal{J}(\thetaa,\phia,\thetav,\phiv)
+
\mathcal{R}(\thetaa,\phia,\phiv,\psi).
\end{align}
\reffig{ModelOverview} shows the overview of the proposed model.

\subsection{Generation processes}

Given the acoustic feature sequence $\x$ of input speech
and a target face image $\y$, 
$\x$ can be converted via
\begin{align}
\hat{\x} = \vmu_{\thetaa}(\vmu_{\phia}(\x), \vmu_{\phiv}(\y)).
\end{align}
A time-domain signal can then be generated using an appropriate vocoder. 
We can also generate a face image corresponding to the input speech $\x$ via
\begin{align}
\hat{\y} = \vmu_{\thetav}(\vmu_{\psi}(\x)).
\end{align}

\subsection{Network architectures}

\begin{figure}[t!]
\centering
  \centerline{\includegraphics[width=.96\linewidth]{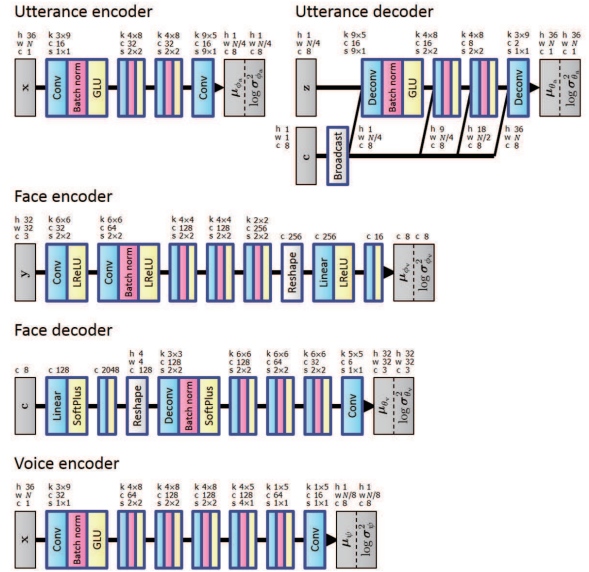}}
  \vspace{-0ex}
  \caption{\footnotesize Architectures of the utterance encoder/decoder, the face encoder/decoder and the voice encoder. Here, the input and output of each of the networks are interpreted as images, where ``h'', ``w'' and ``c'' denote the height, width and channel number, respectively. ``Conv'', ``Deconv'' and ``Linear'' denote convolution, transposed convolution and affine transform,
``Batch norm'', ``GLU'', ``LReLU'' and ``SoftPlus'' denote
batch normalization, GLU, leaky rectified linear unit and softplus layers,
and
``Broadcast'' and ``Reshape'' denote
broadcasting and reshaping operations, respectively. ``k'', ``c'' and ``s'' denote the kernel size, output channel number and stride size of a convolution layer.  }
\label{fig:netarch}
\end{figure}

\noindent
{\bf Utterance encoder/decoder:}~
As detailed in \reffig{netarch},
the utterance encoder/decoder networks are designed using 
fully convolutional architectures with gated linear units (GLUs) \cite{Dauphin2017}. 
The output of the GLU block used in the present model is defined as 
$\mathsf{GLU}(\Vec{X})=\mathsf{B}_1(\mathsf{L}_1(\Vec{X})) \odot \sigma(\mathsf{B}_2(\mathsf{L}_2(\Vec{X})))$ 
where $\Vec{X}$ is the layer input, $\mathsf{L}_1$ and $\mathsf{L}_2$ denote convolution layers, 
$\mathsf{B}_1$ and $\mathsf{B}_2$ denote batch normalization layers, and $\sigma$ denotes a sigmoid gate function.
We used 2D convolutions to design the convolution layers in the encoder and decoder, 
where ${\x}$ is treated as an image of size $D \times N$ with 1 channel. 

\noindent
{\bf Face encoder/decoder:}~
The face encoder/decoder networks are designed using architectures 
inspired by those introduced in \cite{Yan2016} for conditional image generation.

\noindent
{\bf Voice encoder:}~
As with the utterance encoder/decoder, the voice encoder is designed using 
a fully convolutional architecture with GLUs.
As shown in \reffig{netarch}, the voice encoder is designed to produce 
a time sequence of the means (and variances) of latent vectors. 
Here, we expect each of these latent vectors to represent 
information about the voice characteristics of input speech 
within a different time region, which must be time-independent. 
One way of implementing \refeq{r_psi} would be to add a pooling layer 
after the final layer 
so that the network produces the time average of the latent vectors.
However, 
rather than the time average of these values, 
we would want each of these values to be as close to $\c$ as possible.
Hence, here we choose to implement \refeq{r_psi} by treating $\c$ as
a broadcast version of the latent code generated from the face encoder
so that the $\c$ and $\vmu_{\psi}(\x)$ arrays have compatible shapes.

\section{Experiments}

To evaluate the proposed method, we created a 
virtual dataset consisting of speech and face pairs
by combining the Voice Conversion Challenge 2018 (VCC2018) \cite{Lorenzo-Trueba2018} and 
Large-scale CelebFaces Attributes (CelebA) \cite{Liu2015} datasets. 
First, we divided the speech data in the VCC2018 dataset and the face image data in the CelebA dataset
into training and test sets.
For each set, we segmented the speech and face image data 
according to gender (male/female) and age (young/aged) attributes. 
We then treated each pair, which consisted of a speech signal and a face image 
randomly selected from groups with the same attributes, as virtually paired data. 
This indicates that 
the correlation between each speech and face image data pair was artificial.
However, despite this, we believe that testing with this dataset can still provide a useful insight into 
the ability of the present method to capture and leverage the underlying correlation
to convert speech or to generate images in a crossmodal manner.

All the face images were downsampled to 32$\times$32 pixels and
all the speech signals were sampled at 22,050 Hz. 
For each utterance, a spectral envelope,
a logarithmic fundamental frequency (log $F_0$), and
aperiodicities (APs) were extracted every 5 ms using the
WORLD analyzer \cite{Morise2016,pyworld_url}. 
36 mel-cepstral coefficients (MCCs) were then extracted from 
each spectral envelope using the Speech Processing Toolkit (SPTK) \cite{pysptk_url}. 
The aperiodicities were used directly without modification. 
The signals of the converted speech were obtained 
from the converted acoustic feature sequences using the WORLD synthesizer.

We implemented two methods as baselines for comparison, 
which assume the availability of the gender and age attribute label assigned to each data.
One is a naive method that simply adjusts the mean and variance of the 
feature vectors of the input speech for each feature dimension so that they 
match those of the training examples with the same attributes as the input speech. 
We refer to this method as ``Baseline1''.
The other is a two-stage method, which performs face attribute detection followed by attribute-conditioned VC.
For the face attribute detector, we used the same architecture as the face encoder described 
in \reffig{netarch} with the only difference being that we added a softmax layer after the final layer 
so that the network produced the probabilities of the input face image being ``male'' and  ``young''. 
We trained this network using gender/age attribute labels. 
For the attribute-conditioned VC, we used the ACVAE-VC \cite{Kameoka2018b}, also trained using gender/age attribute labels.
We refer to this method as ``Baseline2''.

\begin{figure}[t!]
\centering
  \centerline{\includegraphics[width=.8\linewidth]{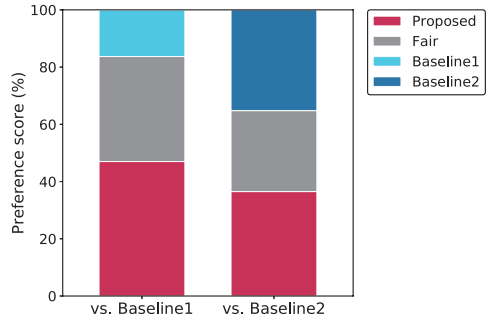}}
  \vspace{-0ex}
  \caption{Results of the ABX test.}
\label{fig:ABXtest}
\end{figure}

We conducted ABX tests to compare how well the voice of speech generated by each of the methods
matched the face image input, where ``A'' and ``B'' were converted speech samples obtained with the proposed and baseline methods
and ``X'' was the face image used for the auxiliary input. 
With these listening tests, ``A'' and ``B'' were presented in random order to eliminate bias in the order of
stimuli. Eleven listeners participated in our listening tests.
Each listener was presented {``A'',``B'',``X''}$\times$ 30 utterances. 
Each listener was then asked to select ``A'', ``B'' or ``fair'' by evaluating which of the two matches ``X'' better.
The results are shown in \reffig{ABXtest}. 
As the results reveal, the proposed method significantly
outperformed Baseline1 and performed comparably to Baseline2.
It is particularly noteworthy that the performance of 
the proposed method was comparable to that of Baseline2 
even though the baseline methods had the advantage of using the attribute labels.
Audio examples are provided at \cite{crossmodalvc_audiodemo_url}.

\begin{figure}[t!]
\centering
  \centerline{\includegraphics[width=.9\linewidth]{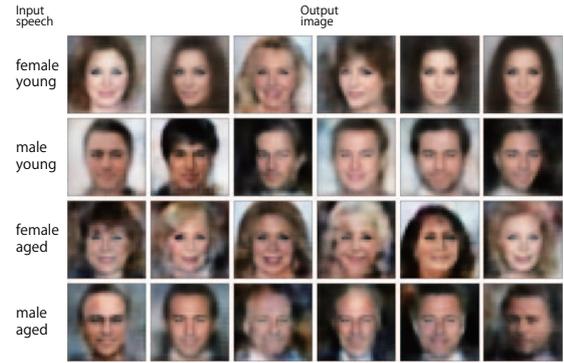}}
  \vspace{-0ex}
  \caption{Examples of the face images predicted from young female,  
  young male, aged female and aged male speech, respectively (from top to bottom rows).}
\label{fig:reconstruct}
 \vspace{-2.0ex}
\end{figure}

\reffig{reconstruct} shows several examples of the 
face images predicted by the proposed method from female and male speech.
As can be seen from these examples, the gender and age of 
the predicted face images are reasonably 
consistent with those of the input speech, demonstrating 
an interesting effect of the proposed method.

\section{Conclusions}

This paper described the first attempt to solve the crossmodal VC problem 
by introducing an extension of our previously proposed non-parallel VC method called ACVAE-VC.
Through experiments using a virtual dataset combining the VCC2018 and CelebA datasets,
we confirmed  
that our method could convert input speech into a voice that matches an auxiliary face image input
and generate a face image that matches input speech reasonably well.
We are also interested in developing a crossmodal text-to-speech system,
where the task is to synthesize speech from text 
with voice characteristics determined by an auxiliary face image input.

\noindent
{\bf Acknowledgements}:
We thank Mr. Ken Shirakawa (Kyoto University) for his help in annotating the virtual corpus
during his summer internship at NTT. 
This work was supported by JSPS KAKENHI 17H01763.

\bibliographystyle{IEEEtran}

\bibliography{Kameoka2019arXiv09}


\end{document}